\documentclass[%
 aip,
rsi,
 amsmath,amssymb,
 reprint,%
]{revtex4-1}

\usepackage{graphicx}
\usepackage{dcolumn}
\usepackage{bm}

\usepackage[utf8]{inputenc}
\usepackage[T1]{fontenc}
\usepackage{mathptmx}
\usepackage{etoolbox}
\usepackage[english]{babel}

\makeatletter
\def\@email#1#2{%
 \endgroup
 \patchcmd{\titleblock@produce}
  {\frontmatter@RRAPformat}
  {\frontmatter@RRAPformat{\produce@RRAP{*#1\href{mailto:#2}{#2}}}\frontmatter@RRAPformat}
  {}{}
}%
\makeatother

\def\n2XPM{n_{2,\mathrm{XPM}}}

\begin{document}


\title[]{Precise measurement of the Kerr coefficient using phase-sensitive pump-probe hyperspectral imaging}

\author{J. K. Wahlstrand}
\author{C. D. Cruz}
 \email{jared.wahlstrand@nist.gov}
\affiliation{ 
$^1$Nanoscale Device Characterization Division, National Institute of Standards and Technology, Gaithersburg, MD 20899, USA
}

\date{\today}

\begin{abstract}
Phase-sensitive pump-probe hyperspectral imaging is a precise technique for absolute two-beam measurements of the optical Kerr coefficient ($n_2$).
The irradiance profile is characterized and background effects are rejected by rastering the pump beam across the probe beam to yield a complex-valued hyperspectral image of the pump-induced nonlinear response.
Information about the temporal irradiance profile is carried in the spectral response.
The technique is demonstrated by measuring $n_2$ of a fused silica sample near 1~$\mu$m wavelength and benchmarked against a measurement using Z-scan [Sheik-Bahae \emph{et al.}, IEEE J. Quantum Electron. \textbf{26}, 760--769 (1990)], the most widely used single-beam technique.
The two measurements are consistent when the two-beam grating effect from the Raman contribution to the nonlinearity is considered.
Uncertainty contributions are described in detail and the outlook is discussed for improvements in precision.
\end{abstract}

\maketitle

\section{Introduction}

The optical Kerr coefficient $n_2$, which characterizes the nonlinear refractive index, encompasses not only the bound electronic third-order nonlinear response that follows optical electric field oscillations on a subfemtosecond timescale, but also the ``nuclear'' or Raman nonlinearity,\cite{hellwarth_origin_1975} including vibrational excitation\cite{stolen_effect_1992,trillo_parametric_1992} and orientational effects,\cite{reichert_temporal_2014} and for sufficiently long pulses or continuous wave lasers, electrostriction\cite{buckland_electrostrictive_1996} and thermal effects.
These additional contributors to the nonlinear response, in particular the Raman response, can impact applications such as fiber communications\cite{stolen_effect_1992} and frequency combs,\cite{cherenkov_raman-kerr_2017} but measurements that distinguish between the various contributors to $n_2$ remain relatively rare outside of well studied materials.
A wide variety of techniques have been developed to measure the nonlinear response.\cite{vermeulen_post-2000_2023}
Single-beam techniques for measuring the Kerr coefficient, such as Z-scan,\cite{sheik-bahae_high-sensitivity_1989} capture the total response.
Isolating the individual components of the effective $n_2$ requires repeated measurements with systematically varied parameters, such as pulse duration or repetition rate.\cite{christodoulides_nonlinear_2010}
This is particularly challenging for the Raman contribution, which accounts for 10~\% to 25~\% of the total nonlinear response in common materials, not much larger than the uncertainty typical of individual single-beam measurements.
Two-beam techniques, such as ultrafast pump-probe measurements, offer a more direct route to resolving time-dependent contributions to $n_2$, including Raman effects.\cite{wahlstrand_absolute_2012,ferdinandus_beam_2013,reichert_temporal_2014,wahlstrand_absolute_2015}

The beam deflection technique\cite{ferdinandus_beam_2013} is the most widely used pump-probe technique for measuring $n_2$, but without extremely careful characterization of beam profiles,\cite{reichert_temporal_2014} it requires referencing to an absolute measurement (typically Z-scan) or to a material with a known response.
Supercontinuum spectral interferometry (SSSI)\cite{kim_single-shot_2002} has been used for absolute (in the sense that $n_2$ can be measured without referencing to a known sample), time-resolved measurements in atmospheric gases,\cite{wahlstrand_absolute_2012,wahlstrand_high_2012,zahedpour_measurement_2015,wahlstrand_absolute_2015,wahlstrand_bound-electron_2018,zahedpour_ultrashort_2019} including measurements of the rotational Raman effect.
While variants of SSSI have been used to characterize the spatiotemporal profile of pulses via the nonlinear phase shift in transparent solids,\cite{hancock_transient-grating_2021} it has not been used to measure $n_2$ in solids beyond initial demonstrations.\cite{kim_single-shot_2002,cruz_phase-sensitive_2024}
There are a few reasons for this.
As developed, SSSI uses collinear, nondegenerate beams, with a pump pulse that can vary widely in wavelength and a probe in the visible range.
While atmospheric gases have nearly negligible dispersion at visible wavelengths, in most solids $n_2$ varies significantly with wavelength, and a measurement at very different pump and probe wavelengths requires modeling based on approximate scaling laws to translate the result to the single-beam $n_2$ value that is most broadly useful.
Pump-probe walk-off due to differing group velocity is an additional complication in nondegenerate pump-probe experiments in solids when the pump and probe wavelengths are widely spaced.\cite{negres_two-photon_2002,negres_experiment_2002}
A primary challenge of any absolute $n_2$ measurement is the need to accurately measure irradiance, and this is compounded in pump-probe techniques because of the need to characterize the overlap between the pump and probe beams.
Finally, to our knowledge pump-probe spectral interferometry has not been previously benchmarked against an established single-beam technique such as Z-scan.

Here, we describe and demonstrate an optimized pump-probe setup for precise measurement of $n_2$ near 1~$\mu$m wavelength.
Instead of visible supercontinuum, we use a tunable optical parametric amplifier (OPA) to produce an infrared probe pulse that is nondegenerate with respect to the pump pulse but is close enough in wavelength that dispersion can be neglected, allowing walk-off free measurements that can be directly compared to single-wavelength measurements.
To improve precision, we adopt a spatially resolved scheme, previously used to measure field ionization in gases,\cite{wahlstrand_bound-electron_2018} to characterize the pump beam irradiance and overlap.
We demonstrate the technique by measuring $n_2$ in fused silica and comparing the result to a Z-scan measurement.

\section{Phase-sensitive pump-probe hyperspectral imaging}

\begin{figure}
    \centering
    \includegraphics[width=8.5 cm]{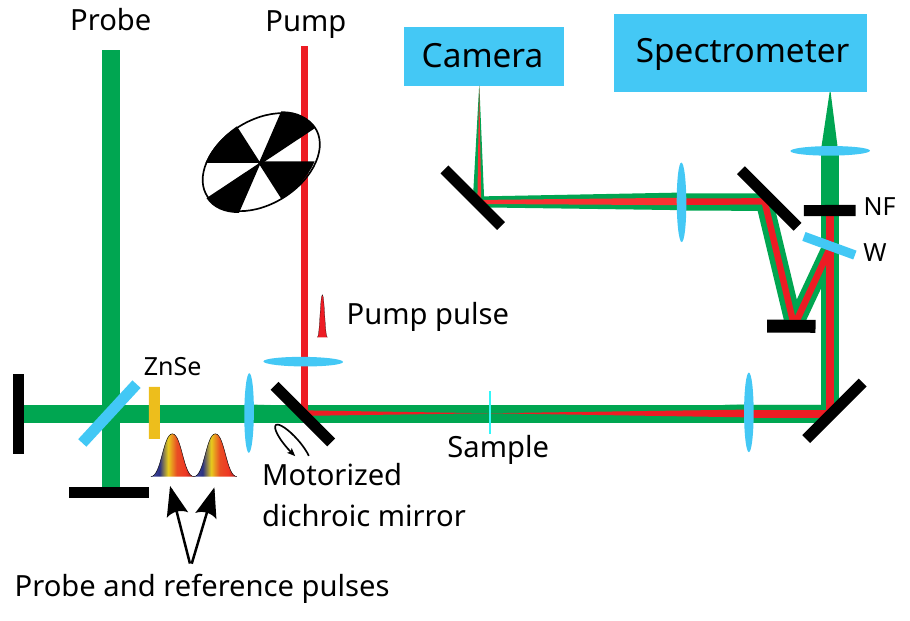}
    \includegraphics[width=8.5 cm]{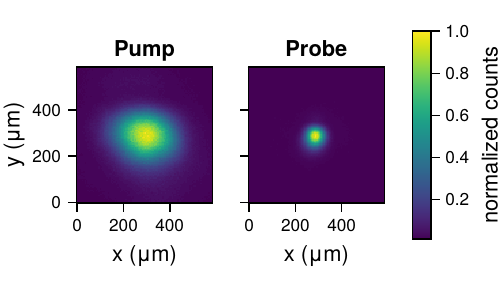}
    \caption{Schematic of the pump-probe setup for measuring 2D phase images. The probe beam is divided into two pulses using a Michelson interferometer. The pump and probe beams are combined and focused such that the probe spot is smaller than the pump spot. A wedge (W) directs a fraction of the pump and probe beams to a relay imaging system (typical images are shown below the diagram) and a motorized mirror scans the pump spot across the probe spot. A notch filter (NF) blocks the pump beam transmitted through the wedge. Probe/reference spectral interferograms are binned by the pump centroid to generate differential phase and amplitude images.}
    \label{setup}
\end{figure}

The pump-probe experiment uses a pump pulse centered at 1024~nm from a Yb laser and a probe pulse centered at 940~nm from an OPA pumped by the same laser.
The bandwidth of the pump spectrum is narrower than the probe spectrum by approximately a factor of 3.
A translation stage in the pump path allows control of the pump pulse delay.
Figure \ref{setup} shows a schematic diagram of the experiment.
The probe beam is split into two pulses, delayed by 6~ps, using a Michelson interferometer, and passes through 4~mm of ZnSe to add chirp. 
The duration of the probe pulse determines the time window over which the nonlinear response can be measured, as will be discussed later.
The pump and probe beams are combined using a dichroic mirror, and each is focused by a lens with a focal length of 600~mm just before the dichroic mirror.
Telescopes in each path (not shown in Fig.~\ref{setup}) shape the beams so that the probe beam is focused to an approximately 75~$\mu$m full width at half maximum (FWHM) waist and the pump beam to an approximately 200~$\mu$m FWHM waist at the common focal plane, where the sample is placed.
A 500~mm focal length lens collimates the transmitted pump and probe beams, and a small fraction is sampled using a glass wedge and relay imaged using a 300~mm focal length lens to a camera.
The magnification of the imaging system is calibrated using a resolution target, backlit with a filtered white light source restricted to 900~nm and 1100~nm to minimize error from chromatic aberration.
The resolution of the imaging system is approximately 30~$\mu$m.
Images of the pump and probe spots at the sample position are shown in Fig.~\ref{setup}.

\begin{figure}
    \centering
    \includegraphics[width=8.5 cm]{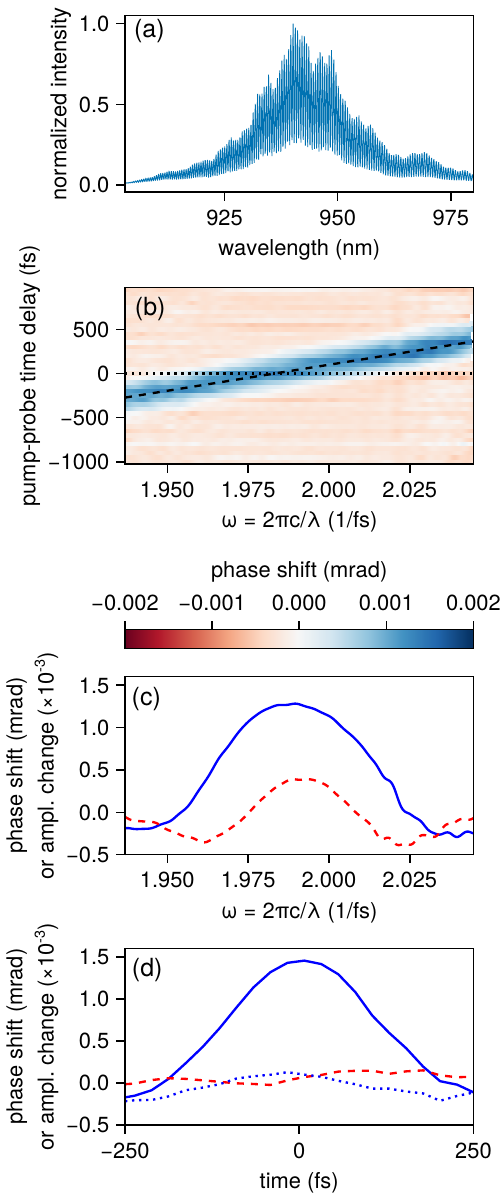}
    \caption{Pump-probe results on fused silica with the pump beam centered on the probe beam position. (a) Probe/reference spectrum, showing spectral interference fringes with a spacing corresponding to the delay of 6~ps. (b) Phase shift as a function of probe frequency and pump-probe time delay. The dashed line indicates the peak phase shift as a function of time delay, which is used to find the probe chirp. (c,d) Phase shift (solid line) and amplitude change (dashed line) at pump-probe delay of zero. (c) Frequency-dependent data. (d) Time-dependent data found by Fourier reconstruction, including the phase shift with no sample due to air (dotted line).}
    \label{expt1}
\end{figure}

Following the wedge, a notch filter blocks the pump beam, and the probe beam is focused onto the entrance slit of a spectrometer.
We record spectra at the laser repetition rate (3~kHz), and a chopper in the pump path blocks two out of every four pump pulses.
The pump-induced, wavelength-dependent phase shift and amplitude change are extracted from spectral interferograms,\cite{tokunaga_femtosecond_1995,kim_single-shot_2002} an example of which is shown in Fig.~\ref{expt1}a.
The fringes arise from interference between the probe and reference pulses.
Analysis of the spectral interferogram allows straightforward extraction of the \emph{frequency-domain} phase and amplitude changes caused by the pump pulse.\cite{tokunaga_femtosecond_1995}
The optical Kerr effect produces a \emph{time-domain} probe phase shift $\Delta \phi(t) = 2\pi\Delta n(t) L/\lambda$, where $\lambda$ is the probe wavelength and $\Delta n(t) \propto n_2 I(t)$ is the refractive index change, with $I(t)$ the time-dependent pump irradiance.
Because the probe pulse is chirped, probe frequency and intrapulse time $t$ are correlated, and for sufficiently slowly varying phase shifts it would be possible to simply map probe frequency to time.
However, a rapidly varying phase shift, as is the case here, causes both a phase shift and an amplitude change in the frequency domain, thereby distorting a simple mapping of frequency to time.
Knowledge of the probe spectral phase allows reconstruction of the time-dependent phase shift,\cite{kim_single-shot_2002} and this capability has been crucial in measurements of the time-dependent nonlinear response of gases.\cite{kim_helium_2002,wahlstrand_absolute_2012,zahedpour_measurement_2015,wahlstrand_absolute_2015,wahlstrand_bound-electron_2018,zahedpour_ultrashort_2019}

The sample discussed here is a fused silica (SiO$_2$) substrate.
Measurement of Fabry-Perot fringes in the transmission spectrum of the sample near 940~nm using the refractive index of fused silica\cite{malitson_interspecimen_1965} $n = 1.4512$ yields a thickness of $L = 511$~$\mu$m~$\pm$~4~$\mu$m (all uncertainty estimates provided here are one standard deviation, $k=1$).
Figure \ref{expt1}b shows the pump-induced phase shift as a function of pump-probe time delay ($\tau$) and probe frequency ($\omega = 2\pi c/\lambda$).
The average power of the pump beam was 177~$\mu$W (measured after the chopper), while the average power of the probe beam was 2~$\mu$W.
We observe a constant phase background of $-250$~$\mu$rad, which might be caused by thermal effects in optical elements or by periodic vibrations and air currents driven by the chopper. A transient phase shift is measured that depends on the pump-probe delay, with a maximum phase shift of 1.8~mrad with respect to the background.
This transient phase shift is caused by the optical Kerr effect.
The spectral phase of the probe pulse $\phi(\omega) = \phi_0 + (\omega-\omega_0)\tau + (1/2)\beta(\omega-\omega_0)^2 + ...$ is characterized by fitting the spectral position of the transient phase shift (dashed line in Fig.~\ref{expt1}b) to $d\phi/d\omega = \tau$.\cite{wahlstrand_optimizing_2016}
We find that to a good approximation, $\tau(\omega) = \beta (\omega-\omega_0)$, with the group delay dispersion parameter $\beta = 1475$~fs$^2$~$\pm$~75~fs$^2$.

Figure \ref{expt1}c shows the probe-frequency-dependent phase shift and amplitude change at a pump-probe time delay of zero (a horizontal slice of the data shown in Fig.~\ref{expt1}b along the dotted line).
Because of the probe chirp, the effect of the pump beam is localized in probe wavelength: for this pump-probe time delay, near 950~nm.
We use the data in Fig.~\ref{expt1}c and $\phi(\omega)$ to extract the time-domain phase shift and amplitude change,\cite{kim_single-shot_2002} shown in Fig.~\ref{expt1}d.
Note that the oscillatory amplitude artifact in the frequency domain (dashed line in Fig.~\ref{expt1}c) is reduced to the noise level in the reconstructed time domain amplitude response (dashed line in Fig.~\ref{expt1}d), and the phase response in the time domain is Gaussian in shape with a FWHM duration of 250~fs.
The data are consistent with a pure phase shift in the time domain, as expected in fused silica for near-infrared pulses of moderate peak irradiance.
The phase shift is expected to follow the pump irradiance envelope in fused silica for pulses longer than 60~fs,\cite{stolen_effect_1992} so in principle we can use the time-dependent phase shift as a measurement of the temporal irradiance profile.\cite{zahedpour_ultrashort_2019}
However, the measured temporal profile scales with the chirp parameter $\beta$.
We therefore also measured the pulse duration with an intensity autocorrelation based on second harmonic generation and found a FWHM pulse duration of 247~fs~$\pm$~$4$~fs, which is in agreement with the observed temporal phase shift and has a smaller uncertainty.

With no sample in place, we measure a background transient phase shift with an amplitude of 300~$\mu$rad, shown as a dotted line in Fig.~\ref{expt1}d, that we attribute to the nonlinear phase shift in the air.
For a 250~fs pulse, $n_2$ in air encompasses both the bound electronic and rotational nonlinearities and is approximately $3 \times 10^{-19}$~cm$^2$/W.\cite{wahlstrand_absolute_2012,zahedpour_measurement_2015}
An effective interaction length of 7~cm, which is consistent with the confocal parameters of the beams, produces a phase shift congruent with our observation.

Pulse energy and duration are two components of the spatiotemporal irradiance profile; the last is the spatial profile.
In our previous reflectivity measurement on silicon,\cite{cruz_phase-sensitive_2024} we characterized the pump beam fluence by imaging the reflected pump spot, and we estimated an uncertainty of 25~\% and an uncertainty due to potentially imperfect pump-probe spatial overlap of 10~\%.
Here we improve the fluence measurement and ensure optimal spatial overlap by adopting 2D+1 SSSI.\cite{wahlstrand_bound-electron_2018}
Here ``2D'' refers to 2 spatial dimensions and ``+1'' refers to the temporal dimension.
In the original implementation, a 1D scanning mirror after the sample was used to build a 2D+1 hyperspectral data set from 1D+1 slices captured by an imaging spectrometer.
In this experiment we use a probe spot smaller than the pump spot, because the acquisition rate is highest for one-dimensional spectra.
As a result we use a 2D scanning mirror before the sample to build a 2D+1 hyperspectral data set from individual spectra.
Recording the pump-probe phase shift as a function of lateral offset between the pump and probe spots reduces the uncertainty in the pump irradiance sampled by the probe beam and ensures that we capture the phase shift at the position of peak pump irradiance.
A counter divides the laser repetition rate by 362, producing a trigger for the imaging camera at 8.3~Hz.
This rate was chosen so that alternating images have the pump present or not present, which allows us to subtract the probe spot from the images.
As spectra are recorded, the imaging camera records pump and probe images at the sample plane.
The centroid $(x_i, y_i)$ of the pump spot is calculated from each image.
The scanning mirror moves the pump spot in a raster pattern and spectra are binned by the measured centroid to create a 2D set of spectral interferograms, from which we create hyperspectral images of the differential phase $\Delta \phi(\lambda,x,y)$ and amplitude $\Delta A(\lambda,x,y)$.
As described earlier, we use the known probe beam spectral phase to reconstruct $\Delta \phi(t,x,y)$ and $\Delta A(t,x,y)$.

\begin{figure}
    \centering
    \includegraphics[width=8.5 cm]{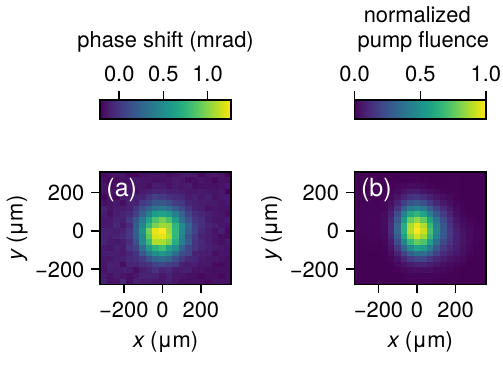}
    \caption{(a) Phase image at $\lambda = 947$~nm. (b) Raw pump spot image. }
    \label{expt2}
\end{figure}

The phase image using $20\times 20$ spatial bins is shown in Fig.~\ref{expt2}a at $\lambda = 947$~nm.
A reconstructed image of the pump spot, created by sampling the pump images at the location of the probe spot, is shown in Fig.~\ref{expt2}b.
A fit of the phase image to a 2D Gaussian, allowing an elliptical beam shape, yields full widths at half maximum (FWHM) of 194.7~$\mu$m~$\pm$~0.8~$\mu$m and 208.2~$\mu$m~$\pm$~0.9~$\mu$m, consistent with the reconstructed pump spot image and individual images of the pump spot.
Given that the phase shift is proportional to the pump fluence, the agreement between the phase image and the camera images verifies that the imaging system is well aligned.
Assuming a Gaussian pulse with irradiance profile $\propto \exp(-t^2/w_t^2)$, where $w_t$ is a temporal width parameter, and a Gaussian spatial distribution $\propto \exp(-x^2/w_x^2-y^2/w_y^2)$, where $w_x$ and $w_y$ are spatial width parameters, the peak irradiance is $I_{\mathrm{peak}} = E_{\mathrm{pulse}}/(\pi^{3/2} w_x w_y w_t)$, where $E_{\mathrm{pulse}}$ is the pulse energy (average power divided by repetition rate).
Using the measured average power, the laser repetition rate, the fluence profile, and the pulse duration previously discussed, we calculate a peak irradiance incident within the sample of $1.01 \times 10^9$~W/cm$^2$~$\pm$~$0.09 \times 10^9$~W/cm$^2$.
The uncertainty shall be discussed in detail later.
Note that the peak fluence calculated assuming a Gaussian distribution and the peak fluence calculated by directly integrating the camera image agreed within a few percent, so to a good approximation we can assume a Gaussian profile here.
We do so only to simplify the uncertainty discussion later.

Each pixel in Fig.~\ref{expt2}a is the average of a few hundred intervals, each containing 362 consecutive spectra.
We calculate the uncertainty in the measured phase shift using the standard deviation of these measurements.
We subtract the peak air background phase shift 306~$\mu$rad~$\pm$~23~$\mu$rad from the peak phase shift measured with the sample, 1602~$\mu$rad~$\pm$~25~$\mu$rad.
Because of Fresnel loss, the presence of the sample reduces the power after the sample by 7~\%, so we reduce the air background to 296~$\mu$rad to account for this loss.
The corrected peak phase shift from the sample is 1306~$\mu$rad~$\pm$~34~$\mu$rad.
Assuming a spatial irradiance profile that does not change as the beam propagates through the sample, which is a very good approximation given our beam parameters, the accumulated peak Kerr phase shift is $\Delta \phi_{\mathrm{peak}} = 2\pi \Delta n_{\mathrm{peak}} L/\lambda$, with $\Delta n_{\mathrm{peak}} = \n2XPM I_{\mathrm{peak}}$, where $\n2XPM$ is the Kerr coefficient for cross phase modulation, which generally differs from the coefficient for self phase modulation.\cite{boyd_nonlinear_2008}
Using
\begin{equation}
\n2XPM = \frac{\Delta \phi_{\mathrm{peak}} \lambda_{\mathrm{probe}}}{2\pi I_{\mathrm{peak}} L},
\label{n2XPM}
\end{equation}
we find $\n2XPM = 3.80\times 10^{-16}$~cm$^2$/W~$\pm$~$0.35\times 10^{-16}$~cm$^2$/W.

\section{Z-scan}

We performed a Z-scan measurement \cite{sheik-bahae_high-sensitivity_1989} using the 1024 nm beam used for the pump beam in the pump-probe measurement.
We found in the pump-probe experiment that the irradiance profile at the focus was to a good approximation Gaussian, but conventional Z-scan relies on propagation as a Gaussian mode, a more stringent condition that the laser beam does not satisfy to a sufficient degree.
We therefore used the Z-scan variant employing a top hat beam.\cite{zhao_zscan_1993}
We expanded the beam using a 4:1 telescope and placed an iris with an aperture of approximately 2.5~mm diameter immediately before a 125~mm focal length lens that focused the beam on the sample.
The beam at the focus was imaged using a camera and is well fit by an Airy spot profile $F(r) = |w_r J_1(r/w_r)/r|^2$, where $J_1$ is the Bessel function of first order, with a characteristic width $w_r=18.4$~$\mu$m~$\pm$~$0.5$~$\mu$m.
The peak fluence $F_{\mathrm{peak}}$ of the Airy spot is found using $E_{\mathrm{pulse}} = \int 2\pi r F(r) dr = 4\pi w_r^2 F_{\mathrm{peak}}$.
The repetition rate of the laser was 10~kHz, and the average power was 14.5~mW~$\pm$~0.7~mW.
Again assuming a Gaussian pulse, the peak irradiance is thus $I_{\mathrm{peak}} = E_{\mathrm{pulse}}/(4\pi^{3/2} w_r^2 w_t)$.
Accounting for the Fresnel reflection at the front surface, we find a peak irradiance within the sample of $1.22 \times 10^{11}$~W/cm$^2 \pm 0.06 \times 10^{11}$~W/cm$^2$.

\begin{figure}
    \centering
    \includegraphics[width=8.5 cm]{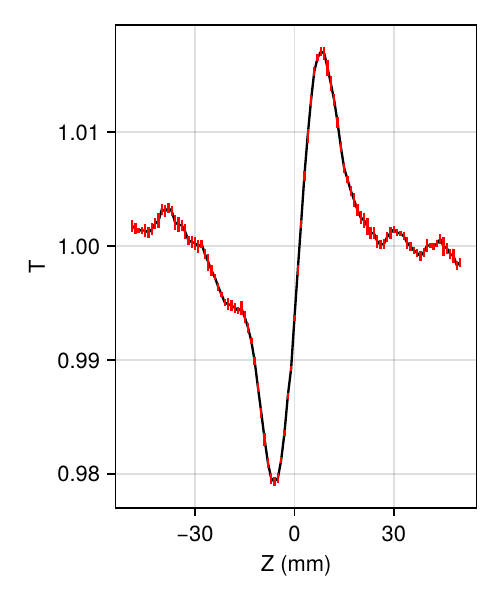}
    \caption{Z-scan experimental results. Closed aperture ($S=0.34$) transmission $T$ as a function of sample position $Z$. Error bars in red show uncertainty calculated from five consecutive scans.}
    \label{Zscan}
\end{figure}

Closed aperture Z-scan data is shown in Fig.~\ref{Zscan}.
The peak-to-valley change in the normalized transmission coefficient is $T_{pv} = 0.038 \pm 0.001$ for an aperture passing a fraction $S = 0.34$ of the beam.
Using $\Delta \phi_Z = 2.7 \tanh^{-1}(T_{pv}/[2.8(1-S)^{1.14}])$,\cite{zhao_zscan_1993} and
\begin{equation}
n_2 = 2.7 \tanh^{-1}\left(\frac{T_{pv}}{2.8(1-S)^{1.14}} \right) \frac{\lambda}{2\pi I L},
\label{n2}
\end{equation}
where $I$ is the time-averaged irradiance, we find $n_2 = 2.18 \times 10^{-16}$~cm$^2$/W~$\pm$~$0.14 \times 10^{-16}$~cm$^2$/W.
Note that we used $I = I_{\mathrm{peak}}/\sqrt{2}$, which assumes a Gaussian pulse and an instantaneous nonlinearity.\cite{sheik-bahae_sensitive_1990}

\section{Analysis and Discussion}
Our Z-scan measurement of $n_2$ in fused silica is consistent with measurements reported in the literature,\cite{desalvo_infrared_1996,milam_review_1998} including recent measurements using a variety of techniques.\cite{flom_ultrafast_2015,kabacinski_nonlinear_2019,jansonas_interferometric_2022,schiek_nonlinear_2023}
However, some caution is advised, since previous work has found that $n_2$ can vary in fused silica from sample to sample.\cite{santran_precise_2004}
For our pump-probe measurement we report the Kerr coefficient for cross-phase modulation $\n2XPM = R n_2$, where $R$ is a ratio that depends on details of the nonlinearity and the wavelengths used.
When translating pump-probe measurements to the single-beam nonlinear coefficient, two-beam interference effects\cite{van_stryland_weak-wave_1982,smirl_picosecond_1982,wahlstrand_effect_2013,wahlstrand_absolute_2015} must be accounted for, which requires a detailed understanding of the nonlinearity's time dependence along with the temporal properties of the laser pulses.
For the bound electronic nonlinearity, where the nonlinear polarization follows the optical field, and generally when degenerate, co-polarized beams are used, $R=2$.\cite{boyd_nonlinear_2008,wahlstrand_effect_2013}
If we apply it here, we find $n_2 = \n2XPM/2 = 1.90 \times 10^{-16}$~cm$^2$/W~$\pm$~$0.17 \times 10^{-16}$~cm$^2$/W, 13~\% lower than the Z-scan measurement.

The situation is more complicated for nondegenerate beams.\cite{wahlstrand_absolute_2015}
In fused silica, the Raman (or ``nuclear'') response has been found by various methods to account for 13~\% to 25~\% of the nonlinearity for long pulses,\cite{hellwarth_origin_1975,heiman_raman_1979,stolen_low-frequency_1982, stolen_development_1984, stolen_raman_1989,smolorz_measurement_1999,schiek_nonlinear_2023} with a complicated response function arising from the characteristic frequencies of the Raman-active phonon modes.
In nondegenerate pump-probe measurements, resonant two-beam effects occur when the pump-probe difference frequency is near that of a vibrational mode.\cite{wahlstrand_absolute_2015}
The two-beam modification of $\n2XPM$ due to the vibrational grating is positive below the resonance and negative above it.\cite{wahlstrand_absolute_2015}
In our measurement, the separation of pump and probe frequencies spans from about 600~cm$^{-1}$ to 1200~cm$^{-1}$, depending on the probe wavelength.
This entire range lies above the main feature in the Raman spectrum of fused silica, which spans 300~cm$^{-1}$ to 500~cm$^{-1}$.\cite{stolen_effect_1992,trillo_parametric_1992}
Consequently, we expect the vibrational contribution in our measurement to be relatively small and negative.
Writing $n_2 = n_{2,e} + n_{2,R}$, where $n_{2,e}$ is the bound electronic contribution and $n_{2,R} = f n_2$ is the Raman contribution with $f$ the Raman fraction, and neglecting the grating two-beam enhancement\cite{wahlstrand_effect_2013} of the Raman response entirely, we obtain $\n2XPM = 2n_{2,e} + n_{2,R} = (2-f) n_2$.

If we use $f = 0.18$, which was reported by Stolen \emph{et al.}\cite{stolen_raman_1989} and is near the center of the range of reported values,\cite{hellwarth_origin_1975,heiman_raman_1979,stolen_low-frequency_1982, stolen_development_1984,stolen_raman_1989,smolorz_measurement_1999,schiek_nonlinear_2023} we find $n_2 = \n2XPM/1.82 = 2.09 \times 10^{-16}$~cm$^2$/W~$\pm$~$0.19 \times 10^{-16}$~cm$^2$/W, which is in closer agreement with our Z-scan measurement.
In the next section we argue that the uncertainty of $R$ is smaller than that of the absolute $n_2$ measurements due to correlations among uncertainty components.

\section{Uncertainty}

Sources of uncertainty are summarized in Table I.
For both pump-probe and Z-scan, average power was measured using the same thermopile power meter, which has a manufacturer-specified calibration uncertainty of 5~\% and linearity of 0.5~\%.
This is the largest single source of uncertainty in our $n_2$ measurements.
An additional uncertainty of a few $\mu$W comes from fluctuations in the measured value; this is negligible in the fused silica measurements presented here, but becomes important for samples that require lower power.
Temporal characterization of the pump pulse (discussed previously) also contributes to both measurements; the temporal characterization of the probe pulse weakly affects the peak phase shift, but not enough to significantly contribute to $\n2XPM$ uncertainty.
Finally, uncertainty in the sample thickness contributes to both measurements; for this sample it is less than 1~\%.

\begin{table}[t]
\begin{tabular}{c | c | c}
Component & Uncertainty (\%) & Contributes to \\
\hline 
Power meter calibration & 5 & Both (correlated) \\
Power meter linearity & 0.5 & Both \\
Pulse duration & 3.2 & Both (correlated) \\
Sample thickness & 0.8 & Both (correlated) \\
Imaging system calibration & 6 & Pump-probe \\
Pump spot image & 0.6 & Pump-probe \\
Peak phase shift & 2.1 & Pump-probe \\
Z-trace & 3.7 & Z-scan \\
Spot image & 1 & Z-scan
\end{tabular}
\caption{Uncertainty budget for the pump-probe and Z-scan measurements of $n_2$ in a fused silica sample.}
\end{table}

An important uncertainty source in the pump-probe experiment is the calibration of the imaging system, as the irradiance scales with this calibration squared.
We calibrated the imaging system using a resolution target and by recording images while translating a knife edge.
The magnification calculated from the calibration agrees with the ratio of focal lengths in the relay imaging system, as expected.
We judge the uncertainty in the calibration to be 3~\%.
The measured phase shift in the pump-probe experiment has an uncertainty of 2.1~\%, characterized using the standard deviation of the data contributing to each spatial bin.
We used standard error propagation using the input uncertainties and Eq.~(\ref{n2XPM}) to find the uncertainty in $\n2XPM$.

In the Z-scan measurement, we calculate uncertainty in $n_2$ based on variables in Eq.~(\ref{n2}): the fraction of the power passing through the iris $S$, which has $<1$~\% uncertainty; $I$, with contributions from $w_r$ (calculated from the camera image), $w_t$, and the average power; $T_{pv}$, the uncertainty of which was found from five Z-scans; and the sample thickness $L$.
Here, the beam profile was measured by attenuating the beam and placing a camera at the focus, so there is no imaging system to calibrate.
We performed Z-scans multiple times, misaligning and realigning it each time, and found that our measurement agreed within $0.1\times 10^{-16}$~cm$^2$/W, suggesting that our uncertainty estimate is reasonable.

Combining Eqs.~(\ref{n2XPM}) and (\ref{n2}) and expressing the irradiances for each experiment in terms of their components, the ratio of the cross-phase modulation response to the self-phase modulation response is
\begin{equation}
R = \frac{\n2XPM}{n_2} = \frac{\Delta \phi_{\mathrm{peak}} w_x w_y \lambda_{\mathrm{probe}} E_{\mathrm{pulse,Z}}}{4\sqrt{2} \Delta \phi_Z w_r^2 \lambda_{\mathrm{pump}}E_{\mathrm{pulse,pp}}},
\end{equation}
where $E_{\mathrm{pulse,Z}}$ is the pulse energy for the Z-scan measurement and $E_{\mathrm{pulse, pp}}$ is the pulse energy for the pump-probe measurement.
We note that $L$ and $w_t$ have dropped out.
Because the same power meter was used for both measurements, we do not include the calibration uncertainty of 5~\%, though we retain an uncertainty of 0.5~\% to account for potential nonlinearity in the power meter response.
We find $R = 1.75$~$\pm$~0.11, an uncertainty of 6.5~\%.
The value $R=2$ lies well outside the range of likely values, confirming a sizable Raman contribution.
The theoretical value of 1.82 discussed earlier, calculated assuming an 18~\% Raman fraction and neglecting the vibrational grating contribution, falls within this uncertainty but on the high side, suggesting a negative Raman grating contribution for our chosen pump and probe wavelengths, which is consistent with previous work on the Raman effect in fused silica.\cite{stolen_effect_1992,trillo_parametric_1992}
A measurement using a more highly chirped probe pulse that includes wavelengths closer to the pump wavelength, to resonantly probe the Raman modes, should be able to measure the Raman component directly via the wavelength-dependent grating response.\cite{wahlstrand_absolute_2015}

There are many potential ways to improve on our implementations of both techniques.
The imaging system calibration is responsible for a higher uncertainty in the pump-probe measurement, and that could be eliminated in a future implementation by placing a camera directly in the focus to verify the pump irradiance profile, as was done in the Z-scan setup.
Here that was not possible because of space constraints in the pump-probe setup.
More care could be put into air background subtraction, and this is essential for thinner samples.
Using a tighter focus tends to reduce the air response because it reduces the Rayleigh length of the beams and thus the effective interaction length.
Placing the sample in a vacuum chamber or a purge cell filled with helium could potentially remove the atmospheric background, or at least reduce it by orders of magnitude.
The assumption of a perfect top hat beam mode and a perfect Gaussian pulse is a potential source of systematic error in the Z-scan setup.
The pump-probe technique also requires some assumptions, but they are in many ways easier to satisfy.
First, we need to assume that the pump beam profile does not change as the beam propagates through the sample, which requires that the Rayleigh length be much larger than $L$, which is well satisfied in the implementation described here.
Furthermore, we assume that the pump-induced phase shift is small enough that the pump beam propagation is not distorted by the Kerr lens.
We can satisfy this by keeping the nonlinear phase shift below $<50$~mrad, which is easy given our sensitivity.
Finally, the pulse shape is often a significant source of uncertainty in Kerr coefficient measurements, and pulse chirp can alter two-beam grating effects.\cite{dogariu_purely_1997,wahlstrand_effect_2013}
Here we used an autocorrelation, which we think is sufficient for our 1024~nm light source, but this does not provide phase information and likely contributes to systematic uncertainty.
Characterizing the pulse shape using frequency-resolved optical gating (FROG)\cite{trebino_measuring_1997} would improve our knowledge of the pulse shape in both experiments.
More sophisticated pulse characterization will be essential when using an OPA to generate the pump pulse.

\section{Conclusion}

We have shown that phase-resolved pump-probe hyperspectral imaging can measure $n_2$ with uncertainty performance comparable to the Z-scan technique.
Scanning the pump beam across the probe beam to build a spatiotemporal image of the nonlinear phase shift enables precise characterization of the irradiance profile, the largest source of uncertainty in $n_2$ measurements.
The pump-probe technique offers higher sensitivity than our Z-scan measurement, though we note that more sensitive implementations of Z-scan have been developed.\cite{xia_eclipsing_1994}
In nondegenerate pump-probe measurements, two-beam grating effects from cumulative processes such as free-carrier excitation and thermal effects are suppressed when there is no overlap between the spectra of the pump and probe pulses.\cite{wahlstrand_effect_2013}
This is an advantage over degenerate pump-probe measurements, which require extensive theoretical analysis to account for two-beam grating effects.\cite{smirl_picosecond_1982,wahlstrand_effect_2013,furey_im_2021}
The disadvantage of nondegenerate measurements relative to single-beam measurements is that the $n_2$ measured can differ from the single-beam $n_2$, both due to dispersion in $n_2$ and, as was the case here, because time-dependent contributions such as the Raman response are modified by grating effects.
We emphasize that these grating effects can be accounted for if the time dependence of the nonlinearity is known.
By using a nondegenerate probe beam with a wavelength relatively close to the pump beam wavelength, we suppressed most two-beam grating effects while measuring an $n_2$ that is close to the degenerate nonlinearity of greatest interest in applications.

In H$_2$ and D$_2$ gas, the high vibrational frequency made it straightforward to use nondegenerate pump-probe spectroscopy to precisely extract the Raman contribution to $n_2$.\cite{wahlstrand_absolute_2015}
The same approach could in principle be applied to solid samples; here it was not possible due to the limited tuning range of our OPA and the relatively low frequency of the most important phonon band in fused silica.
Use of a higher probe chirp would help to disentangle the frequency dependence of the Raman response from the effect of the pump pulse duration.
A polarization-dependent measurement would also be informative, as the Raman response is often strongly polarization dependent.\cite{stolen_effect_1992}

Our results demonstrate that 2D+1 spectral interferometry is capable of absolute measurements and may be less susceptible to systematic error than Z-scan, though further work is needed to explore this.
Phase-sensitive pump-probe imaging can be used in situations where the sample is too thin to produce a measurable signal in Z-scan or too absorptive for Z-scan to be applicable.
The technique is compatible with scanning microscopy, making it suitable for small and/or inhomogeneous samples.\cite{fischer_invited_2016}
The complexity and relatively low throughput of 2D+1 spectral interferometry will assuredly prevent it from being deployed outside of specialized laboratories, unlike Z-scan.
It can nevertheless be employed to develop standard reference samples for relative measurements using Z-scan and other simpler methods.
Measurements of one sample relative to another are immune to errors in irradiance profile characterization,\cite{bridges_z-scan_1995} so developing better reference samples could have an outsized impact on reducing uncertainty.
The results presented here demonstrate the technique's promise for quantifying the Raman component of $n_2$, and it is also well suited for precise nondegenerate measurements of the complex third-order nonlinear susceptibility, both in reflection\cite{cruz_phase-sensitive_2024} and transmission geometries.
The precisely characterized irradiance profile will also benefit precise measurements of other irradiance-dependent phenomena, such as free-carrier absorption.

\section{Acknowledgments}
We thank John Stephenson for helpful suggestions and a critical reading of the manuscript.
J.K.W. acknowledges useful discussions with H. M. Milchberg and E. W. Van Stryland.

\section*{Data Availability Statement}

The data that support the findings of this study are available from the corresponding author upon reasonable request.

\section*{Author Declarations} The authors have no conflicts of interest to disclose.

\end{document}